\documentclass[twocolumn,amsfonts,showpacs,superscriptaddress,nofootinbib]{revtex4-1}
\usepackage[dvipdfmx]{graphicx}
\usepackage{amssymb}
\usepackage{bm}
\usepackage{color}

\newcommand{\bc}{\begin{center}}
\newcommand{\ec}{\end{center}}
\def\ba#1{\begin{array}{#1}\displaystyle}
\newcommand{\ea}{\end{array}}

\newcommand{\beq}{\begin{equation}}
\newcommand{\eeq}{\end{equation}}
\newcommand{\beqa}{\begin{eqnarray}}
\newcommand{\eeqa}{\end{eqnarray}}
\newcommand{\no}{\nonumber}
\newcommand{\n}{\nonumber\\}
\newcommand{\bi}{\begin{itemize}}
\newcommand{\ei}{\end{itemize}}

\def\lt#1{\left#1}
\def\rt#1{\right#1}

\def\frc#1#2{\frac{#1}{#2}}

\newcommand{\p}{\partial}

\newcommand{\rtheta}{\overrightarrow{\theta}}
\newcommand{\ltheta}{\overleftarrow{\theta}}
\newcommand{\rdelta}{\overrightarrow{\delta}}

\newcommand{\bra}{\langle}
\newcommand{\ket}{\rangle}

\newcommand{\R}{{\mathbb{R}}}

\newcommand{\Or}{{\cal O}}

\newcommand{\ep}{\epsilon}

\newcommand{\Tr}{{\rm Tr}}

\newcommand{\ri}{{\rm i}}
\newcommand{\dd}{{\rm d}}
\newcommand{\LL}{\mathrm{LL}}
\newcommand{\dr}{\mathrm{dr}}

\newcommand{\qq}{{\tt q}}
\newcommand{\jj}{{\tt j}}

\def\eqref#1{(\ref{#1})}

\begin{document}

\title{Emergent hydrodynamics in integrable quantum systems out of equilibrium}

\author{Olalla A. Castro-Alvaredo}
\affiliation
{Department of Mathematics, City University London, Northampton Square EC1V 0HB, U.K.
}
\author{Benjamin Doyon}
\affiliation
{Department of Mathematics, King's College London, Strand WC2R 2LS, UK
}
\author{Takato Yoshimura}
\affiliation
{Department of Mathematics, King's College London, Strand WC2R 2LS, UK
}


\begin{abstract}
Understanding the general principles underlying strongly interacting quantum states out of equilibrium is one of the most important tasks of current theoretical physics. With experiments accessing the intricate dynamics of many-body quantum systems, it is paramount to develop powerful methods that encode the emergent physics. Up to now, the strong dichotomy observed between integrable and non-integrable evolutions made an overarching theory difficult to build, especially for transport phenomena where space-time profiles are drastically different. We present a novel framework for studying transport in integrable systems: hydrodynamics with infinitely-many conservation laws. This bridges the conceptual gap between integrable and non-integrable quantum dynamics, and gives powerful tools for accurate studies of space-time profiles. We apply it to the description of energy transport between heat baths, and provide a full description of the current-carrying non-equilibrium steady state and the transition regions in a family of models including the Lieb-Liniger model of interacting Bose gases, realized in experiments.
\end{abstract}

\maketitle

\section{Introduction}
Many-body quantum systems out of equilibrium give rise to some of the most important challenges of modern physics \cite{Eisrev}. They have received a lot of attention recently, with experiments on quantum heat flows \cite{Pierre_et_al,Brant13}, generalized thermalization \cite{KWW,LE} and light-cone effects \cite{lightcone}. The leading principle underlying non-equilibrium dynamics is that of local transport carried by conserved currents. Deeper understanding can be gained from studying non-equilibrium, current-carrying steady states, especially those emerging from unitary dynamics \cite{Ruelle}. This principle gives rise to two  seemingly disconnected paradigms for many-body quantum dynamics. On the one hand, taking into account only few conservation laws, emergent hydrodynamics \cite{Qhydro1,Qhydro2,Qhydro3,Qhydro4,hydro} offers a powerful description where the physics of fluids dominates \cite{Nat-Phys,CKY,BDirre,Pour,rarefact1,rarefact2}. On the other hand, in integrable systems, the infinite number of conservation laws are known to lead to generalized thermalization \cite{GGE,INWCEP,Dtherm} (there are many fundamental works on the subject, see the review \cite{EFreview}), and the presence of quasi-local charges has been shown to influence transport \cite{Drude_bound1,Drude_bound2} (see the review \cite{IPZreview}). However, except at criticality \cite{BD2012,BD-long} (see the review \cite{BDreview}), no general many-body emergent dynamics has been proposed in the integrable case; with the available frameworks, these two paradigms seem difficult to bridge. The study of pre-thermalization or pre-relaxation under small integrability breaking \cite{pretherm1,pretherm2,EFreview,BDreview}, the elusive quantum KAM theorem \cite{qKAM1,qKAM2}, the development of perturbation theory for non-equilibrium states, and the exact treatment of non-equilibrium steady states and of non-homogeneous quantum dynamics in unitary interacting integrable models remain difficult problems.

\begin{figure}
\includegraphics[width=7cm]{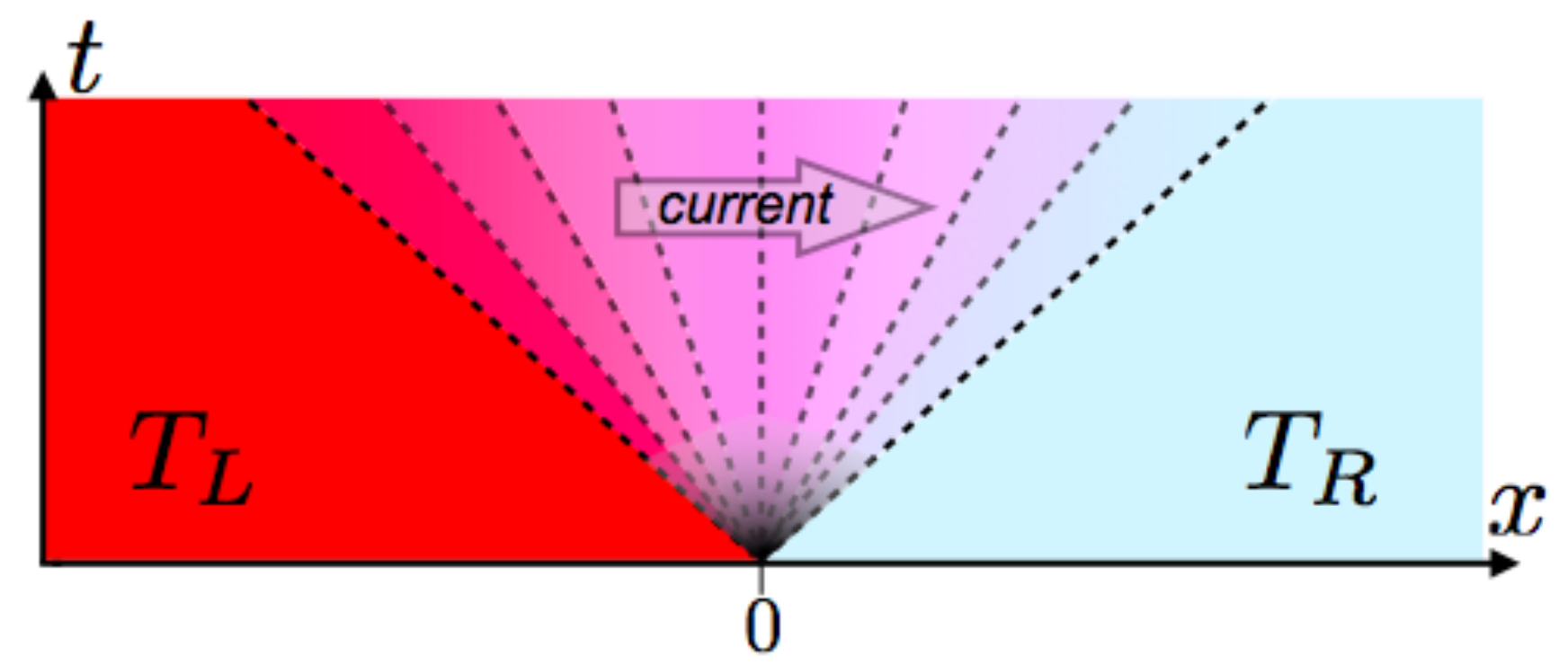}
\caption{The partitioning protocol. With ballistic transport, a current emerges after a transient period. Dotted lines represent different values of $\xi=x/t$. If a maximal velocity exists (e.g. due to the Lieb-Robinson bound), initial reservoirs are unaffected beyond it (light-cone effect). The steady state lies at $\xi=0$.}
\label{figsetup}
\end{figure}

In this paper, using the recent advances on generalized thermalization and developing further aspects of integrability, we propose a solution to such problems by deriving a general theory of hydrodynamics with infinitely-many conservation laws. The theory, applicable to a large integrability class, is derived solely from the fundamental tenet of emerging hydrodynamic: local entropy maximization (often referred to as local thermodynamic equilibrium) \cite{lte1,lte2,lte3,lte4,spobook}.  Focussing on quantum field theory (QFT) in one space dimension, we then study a family of models that include the paradigmatic Lieb-Liniger model \cite{LLmodel} for interacting Bose gases, explicitly realized in experiments \cite{LLexp1,LLexp2,LLexp3,KWW,LE}. We concentrate on far-from-equilibrium states driven by heat baths in the partitioning protocol \cite{spo,Ruelle,BD2012,BD-long} (see Fig. \ref{figsetup}). We provide currents and full space-time profiles which are in principle experimentally accessible, beyond linear response and for arbitrary interaction strengths. We make contact with the physics of rarefaction waves, and with the concept of quasi-particle underlying integrable dynamics.

{\em Note added:} After a first version of this paper appeared as a preprint, similar dynamical equations as those derived here were independently obtained in the integrable XXZ Heisenberg chain by assuming, in addition to local entropy maximization, an underlying kinetic theory \cite{BCDF16}. Solutions to these equations of the same type as those considered here were constructed and confirmed by numerical simulations.

\section{Setup}
Let two semi-infinite halves (which we refer to as the left and right reservoirs) of a homogeneous, short-range one-dimensional quantum system be independently thermalized, say at temperatures $T_L$ and $T_R$. Let this initial state $\bra\cdots\ket_{\rm ini}$ be evolved unitarily with the Hamiltonian $H$ representing the full homogeneous system. One may then investigate the steady state that occurs at large times (see e.g. \cite{BDreview}),
\beq\label{stalim}
	{\tt O}^{\rm sta}  := \lim_{t\to\infty} \bra e^{\ri Ht} O e^{-\ri H t}\ket_{\rm ini},\quad O\mbox{ local observable.}
\eeq
If the limit exists, it is a maximal-entropy steady state involving, in principle, all (quasi-)local conserved charges of the dynamics $H$ (see \eqref{dens} below). Generically, the dynamics only admits a single conserved quantity, $H$ itself: this means that, due to diffusive processes, ordinary Gibbs thermalization occurs. However, when conserved charges exist that are odd under time reversal, the steady state may admit nonzero stationary currents. This indicates the presence of ballistic transport, and the emergence of a current-carrying state that is far from equilibrium (breaking time-reversal symmetry). This is the partitioning protocol for building non-equilibrium steady states. See Fig. \ref{figsetup}.

The study of such non-equilibrium steady states has received a large amount of attention recently (see \cite{BDreview} and references therein). They form a uniquely interesting set of states: they are simple enough to be theoretically described, yet encode nontrivial aspects of non-equilibrium physics.  They naturally occur in the universal region near criticality described by QFT, where ballistic transport emerges thanks to continuous translation invariance; and in integrable systems, where it often arises thanks to the infinite family of conservation laws.

Works \cite{BD2012,BD-long} open to door to their study at strong-coupling critical points with unit dynamical exponent, obtaining in particular the full universal time evolution. The steady state was found to be homogeneous within a light cone, with the energy current being
\beq\label{CFT}
	{\tt j}^{\rm sta} = \frc{\pi ck_B^2}{12\hbar} (T_L^2 - T_R^2)
\eeq
where $c$ is the central charge of the conformal field theory (CFT) (below we set $k_B=\hbar=1$). This result arises from the independent thermalization of emerging left- and right-moving energy carriers (chiral separation). It was numerically verified \cite{Moore12} and agrees with recent heat-flow experiments \cite{Pierre_et_al}. It was generalized using hydrodynamic methods to higher-dimensional critical points \cite{Nat-Phys,CKY,rarefact1,rarefact2} and to deviations from criticality \cite{BDirre,Pour,rarefact2}. Under conditions that are fulfilled in universal near-critical regions, inequalities that generalize \eqref{CFT} can be derived \cite{D_2014,BDreview} (here with unit Lieb-Robinson velocity \cite{lr})
\beq\label{ineq}
	\frc{{\tt e}^L - {\tt e}^R}{2} \geq {\tt j}^{\rm sta} \geq
	\frc{{\tt k}^L - {\tt k}^R}{2}
\eeq
where ${\tt e}^{L,R}$ and ${\tt k}^{L,R}$ are, respectively, the energy densities and the pressures in the left and right reservoirs\footnote{The second inequality in \eqref{ineq} was shown in \cite{D_2014}, with the goal of establishing the existence of current-carrying non-equilibrium steady states. The first was suggested to BD by M. J. Bhaseen shortly afterwards, and can be shown in the same manner.}.

Many further results exist in free-particle models (see \cite{BDreview} and references therein), where independent thermalization of right- and left-movers still hold. In contrast, however, only conjectures and approximations are available for interacting integrable models \cite{castroint,DeLucaVitiXXZ,Zotos16}. In addition, a striking dichotomy is observed between integrable situations and hydrodynamic-based results: for instance, conformal hydrodynamics is expected to emerge in strong-coupling CFT \cite{Nat-Phys,CKY}, leading to shock structures, but generically fails in free-particle conformal models \cite{doyonKG}, where transition regions are smooth. This points to the stark effect of integrability on non-equilibrium quantum dynamics, still insufficiently understood with available techniques.

\section{Emerging hydrodynamics in quantum systems}\label{secthydro}
Let us recall the basic concepts underlying the hydrodynamic description of many-body quantum systems, and its use in the setup described above  (similar concepts exist in many-body classical systems).

Let $Q_i,\,i\in\{1,2,\ldots,N\}$ be local conserved quantities in involution. These are integrals of local densities $q_i(x,t)$, and the conservation laws take the form $\p_t q_i(x,t)+\p_x j_i(x,t)=0$, where $j_i$ are the associated local currents. A Gibbs ensemble is a maximal-entropy ensemble under conditions fixing all averaged local conserved densities. It is described by a density matrix
\beq \label{dens}
	\rho = e^{-\sum_i \beta_i Q_i} / \Tr\big[e^{-\sum_i \beta_i Q_i}\big],
\eeq
where $\beta_i$ are the associated potentials. For instance, $Q_1$ is taken as the Hamiltonian, and $\beta_1$ is the inverse temperature. We will denote $\underline\beta=(\beta_1,\beta_2,\ldots,\beta_N)$ the vector representing this state, and $\bra \cdots\ket_{\underline\beta}$ the averages.

Clearly, the Gibbs averages of local densities ${\tt q}_i=\bra q_i\ket_{\underline\beta}$ (these are independent of space and time by homogeneity and stationarity) may be seen as defining a map from states to averages, $\underline\beta\mapsto \underline{{\tt q}}$. This is expected to be a bijection: the set of averages fully determines the set of potentials. Therefore, the current averages ${\tt j}_i=\bra j_i\ket_{\underline\beta}$ are functions of the density averages, 
\beq\label{eos}
	\underline{{\tt j}} = \underline{\cal F}({\underline{\tt q}}).
\eeq
These are the equations of state, and are model-dependent. The averages $\underline{\tt q}$ can be generated by differentiation of the (specific, dimensionless) free energy $f_{\underline{\beta}}$. Similarly, one can show \cite{BDreview} (see Appendix \ref{curpot}) that there exists a function $g_{\underline \beta}$ that, likewise, generates the currents,
\beq\label{fe}
	\underline{\tt q}=\nabla_{\underline\beta}f_{\underline\beta},\quad
	\underline{\tt j}=\nabla_{\underline\beta}g_{\underline\beta}.
\eeq

A hydrodynamics description of quantum dynamics is expected to emerge at large space-time scales. This has been exploited, in the present setup, in \cite{Nat-Phys,CKY,Pour,BDirre,rarefact1,rarefact2}. The emergence of hydrodynamics is solely based on the assumption of local entropy maximization (or local thermodynamic equilibrium\footnote{The phrase ``local thermodynamic equilibrium'' is often used to describe the fluid cells, however it might be slightly misleading as it seems to suppose that fluid cells are in equilibrium; in order to have nontrivial hydrodynamics this however is not the case, as one needs the presence of nonzero ballistic currents, breaking time-reversal symmetry.}). Technically, this is the assumption that averages of local quantities $\bra O(x,t)\ket$ tend uniformly enough, at large times, to averages evaluated in local Gibbs ensembles $\bra O\ket_{\underline\beta(x,t)}$ with space-time dependent potentials $\underline\beta(x,t)$. Physically, this is a consequence of {\em separation of scales}, as follows (see for instance \cite{spobook}).

Assume that, after some time, physical properties vary only on space-time scales that are much larger than microscopic scales. This may be referred to as the ``local relaxation time''. From that time on, microscopic processes such as particle collisions or inter-site interactions give rise to fast, local relaxation: the reaching of a (approximate) steady state on space-time scales small compared to variations but large enough for thermodynamics to be applicable. By Boltzmann's phase-space argument, these local steady states are obtained from entropy maximization, and as usual maximization is under the conditions provided by conservation laws (properties of the microscopic dynamics). That is, on each space-time ``fluid cell'' a Gibbs state is (very nearly) reached. Neighboring Gibbs states are different, but their variations are small. This is local entropy maximization.

Assume local entropy maximization. On each fluid cell, the Gibbs state is initially characterized by the values of the conserved densities at the local-relaxation time. The large-scale dynamics is thereon obtained from conservation laws, as follows. Consider microscopic conservation in integral form, $\int_{x_1}^{x_2} \dd x\, \big(q_i(x,t_2)
	- q_i(x,t_1)\big)
	+
	\int_{t_1}^{t_2} \dd t\,\big(j_i(x_2,t)
	- j_i(x_1,t)\big) = 0$. Since averages of densities and currents, after the local relaxation time, take the form $\bra q_i(x,t)\ket = \bra q_i\ket_{\underline\beta(x,t)}$ and $\bra j_i(x,t)\ket = \bra j_i\ket_{\underline\beta(x,t)}$ uniformly enough, we have
\beq
	\int_{x_1}^{x_2} \dd x\, \big(\underline{\tt q}(x,t_2)
	- \underline{\tt q}(x,t_1)\big)
	+
	\int_{t_1}^{t_2} \dd t\,\big(\underline{\tt j}(x_2,t)
	- \underline{\tt j}(x_1,t)\big) = 0
\eeq
where $\underline{{\tt q}}(x,t)= \bra \underline{q} \ket_{\underline\beta(x,t)}$ and $\underline{{\tt j}}(x,t) = \bra\underline{j} \ket_{\underline\beta(x,t)}$. Here, integrals may be taken to cover a {\em macroscopic number of fluid cells}: these become macroscopic conservation equations. Macroscopic conservation equations can be re-written in differential form, with differentials representing small variations amongst fluid cells:
\beq\label{conshy}
	\p_t \underline{{\tt q}}(x,t) + \p_x \underline{{\tt j}}(x,t) = 0.
\eeq
These are the pure hydrodynamic (Euler-type) equations, representing the slow, large-scale quantum dynamics of conserved densities and currents flowing amongst neighboring cells.

The problem of emergence of hydrodynamics in many-body systems is one of the most important unsolved problem of modern mathematical physics. Although there are few proofs of emergence of hydrodynamics, there is strong evidence for the validity of emerging Euler equations in many situations; see the books \cite{lte1,lte2,lte3,lte4,spobook}, and the recent paper \cite{mendspo} for a study of emerging Euler equations in classical anharmonic chains.

Combined with the equations of state \eqref{eos}, Euler equations \eqref{conshy} give
\beq\label{hydro}
	\p_t \underline{{\tt q}}(x,t) + J(\underline{{\tt q}}(x,t))
	\p_x \underline{{\tt q}}(x,t) = 0
\eeq
where $J(\underline{{\tt q}}) := \nabla_{\underline {\tt q}} \underline{\tt j}$ is an $N$ by $N$ matrix, the Jacobian matrix of the transformation from densities to currents,
\beq\label{J}
	J(\underline{{\tt q}})_{ij} = \p {\cal F}_i(\underline{{\tt q}})/
	\p {\tt q}_j.
\eeq
Equations \eqref{hydro} are the emergent pure hydrodynamic  equations in quasi-linear (or characteristic) form \cite{hydro}.
The complete model dependence, including all quantum effects, is encoded, besides the number $N$ of conserved quantities, in the Jacobian $J(\underline{\tt q})$.

The density averages $\underline{\tt q}$, like the potentials $\underline \beta$, correspond to a set of state coordinates. One may choose any other set of state coordinates $\underline{\tt n}$, with $\underline{\tt q} = \underline{\cal F}^{\tt q}(\underline{\tt n})$ and $\underline{\tt j} = \underline{\cal F}^{\tt j}(\underline{\tt n})$. A similar equation is obtained,
\beq\label{hydron}
	\p_t \underline{{\tt n}}(x,t) + J(\underline{{\tt n}}(x,t))
	\p_x \underline{{\tt n}}(x,t) = 0,
\eeq
where $J(\underline{\tt n}) = \big(\nabla_{\underline{\tt n}} \underline{\tt q}\big)^{-1} \nabla_{\underline{\tt n}} \underline{\tt j}$. Observe that $J(\underline{\tt n})$ and $J(\underline{\tt q})$ are related to each other by a similarity transformation: $J(\underline{\tt n}) = \big(\nabla_{\underline{\tt n}} \underline{\tt j}\big)^{-1} J(\underline{\tt q})|_{\underline{\tt q} = \underline{\cal F}^{\tt q}(\underline{\tt n})} \nabla_{\underline{\tt n}} \underline{\tt j}$. Therefore, the spectrum of $J(\underline{\tt n})$ is independent of the choice of coordinates, and is a fundamental property of the model. We will denote this spectrum by $\{v_i^{\rm eff}(\underline{\tt n}),i=1,\ldots,N\}$.

Choosing coordinates $\underline{\tt n}$ that diagonalize $J(\underline{\tt n})$ one obtains
\beq
	\p_t  {\tt n}_i(x,t) + v_i^{\rm eff}(\underline{\tt n}(x,t)) \p_x {\tt n}_i(x,t) = 0.
\eeq
These express the vanishing of the convective derivatives, representing the constance of each fluid mode ${\tt n}_i(x,t)$ on fluid cells. The eigenvalues $v_i^{\rm eff}(\underline{\tt n}(x,t))$ are therefore interpreted as the {\em propagation velocities} of these normal modes. The normal modes interact with each other only through the propagation velocities, which is generically a function of all state coordinates.

Let us now apply the above to the solution of the partitioning problem. For clarity of the following discussion, we come back to the ${\tt q}$-coordinates (but it is easy to generalize to any coordinates ${\tt n}$). Consider the large-scale limit $(x,t)\mapsto (ax,at),\;a\to\infty$. Because \eqref{hydro} is invariant under this scaling, in the limit, if it exists, the solution is also invariant. Thus we may assume {\em self-similar solutions} $\underline\beta(x,t) = \underline\beta(\xi)$ where $\xi=x/t$, and  \eqref{hydro} becomes an eigenvalue equation,
\beq\label{hydroscale}
	(J(\underline{\tt q})-\xi {\bf 1})\p_\xi \underline{\tt q} = 0.
\eeq
The initial condition is determined by the state at the local relaxation time (at which the fluid-dynamics description starts to be valid). This state is unknown, as it depends on the full quantum dynamics, but its asymptotic at large $|x|$ is identical to that of the original state. In the large-scale solution, the initial condition $t\to0^+$ is implemented as asymptotic conditions as $\xi\to\pm\infty$. Therefore it only depends on the asymptotic form of the initial state, and we impose
\beq\label{hydroini}
	\lim_{\xi\to\pm\infty}\underline{\tt q}(\xi) =
	\lim_{x\to\pm\infty} \bra \underline{q}(x,0)\ket_{\rm ini}.
\eeq
In the present setup, these involve Gibbs states at potentials $\underline{\beta}^{R,L}$:
\beq
	\lim_{x\to\pm\infty} \bra \underline{q}(x,0)\ket_{\rm ini}
	=\bra\underline{q}\ket_{{\underline{\beta}^{R,L}}}
\eeq
and the steady-state averages are given by
\beq
	\underline{\tt q}^{\,\rm sta}:=\underline{\tt q}(\xi=0),\ \ 
	\underline{\tt j}^{\,\rm sta}:=\underline{\tt j}(\xi=0).
\eeq

The solution to the eigenvalue equation \eqref{hydroscale} and initial conditions \eqref{hydroini} provides the exact large-scale asymptotic form of the full quantum solution, along any ray $x=\xi t$ (see Fig. \ref{figsetup}). The eigenvalue equation \eqref{hydroscale} represents the small changes of averages along various rays, due to the exchange of conserved charges amongst fluid cells. The set of eigenvalues of $J({\underline {\tt q}})$ -- the available propagation velocities in the state characterized by the averages $\underline {\tt q}$ -- form a finite, discrete set for finite $N$.

Solutions to \eqref{hydroscale}, \eqref{hydroini} are typically composed of regions of constant $\underline{\tt q}$-values separated by transition regions  \cite{hydro}. Transition regions may be of two types: either shocks (weak solutions), where $\underline{\tt q}$-values display finite jumps, or rarefaction waves, where they form a smooth solution to \eqref{hydroscale}. Rarefaction waves, the most natural type of solution, cannot, generically, cover the full space between two reservoirs. Indeed, \eqref{hydroscale} specifies that the curve traced by the solution in the $\underline{\tt q}$-plane must at all points be tangent to an eigenvector of $J(\underline{\tt q})$. Since eigenvectors -- and available propagation velocities -- form a discrete set, smooth variations of $\underline{\tt q}$ along the curve imply a unique choice of eigenvector at each point (except possibly at points where eigenvalues cross). Thus, the curve is completely determined by its initial point, and cannot join two arbitrary reservoir values. That is, in ordinary pure hydrodynamics, shocks are often required.

\section{Hydrodynamics with infinitely many currents}
In integrable systems, there are infinitely many local conservation laws. In fact, this space is enlarged to that of ``pseudolocal'' conservation laws, where the densities $q_i(x,t)$ and currents $j_i(x,t)$ are supported on extended spacial regions with weight decaying fast enough away from $x$. This enlargement plays an important role in non-equilibrium quantum dynamics \cite{Drude_bound1,Drude_bound2,INWCEP,Dtherm,IPZreview}. Under maximal-entropy principles, Gibbs states are then replaced by generalized Gibbs ensembles (GGE) \cite{GGE,EFreview,Dtherm}: formally the limit $N\to\infty$ of the density matrix \eqref{dens}, involving all basis elements in the space of conserved pseudolocal charges. We choose $Q_1=H$ (the Hamiltonian) and $Q_2=P$ (the momentum operator).

Under the influence of infinitely many conservation laws, the picture of local entropy maximization is still expected to hold: all physical principles underlying it stay unchanged, the only difference being the use of GGEs instead of Gibbs ensembles. This, along with the emergence of self-similar solutions in the partitioning protocol, are our working hypotheses; see Appendix \ref{apparg} for a discussion. The emergence of a generalized type of hydrodynamics was proven in the classical hard-rod problem \cite{dobrods,spobook}, whose relation with the present quantum problem we will study in a future work. The emergence of self-similar solutions was observed numerically in the quantum XXZ chain in \cite{SM13}. In free-particle quantum models, hydrodynamic ideas and related semi-classical approximations, as well as ray-dependent local entropy maximization, were studied in various works, see \cite{AKR08,WCK13,collura1,collura2,VSDH,BF16}.

Looking for a full solution to the infinity of equations \eqref{hydro}, \eqref{hydroscale} and \eqref{hydroini}, an appropriate choice of state variables is crucial. A powerful way is to recast them into the quasi-particle language underlying the thermodynamic Bethe ansatz (TBA) method for integrable systems \cite{tba1}. Using this language, we derive the exact GGE equations of state, and the ensuing {\em generalized hydrodynamics} equation. We determine the exact normal modes and propagation velocities, and obtain full ray-dependent solutions.

\subsection{GGE equations of state}

We assume that the spectrum of stable quasi-particles is composed of a single quasi-particle species of mass $m$ (see Appendix \ref{morepart} for a many-particle generalization). The dispersion relation is encoded via a parametrization $E(\theta)$, $p(\theta)$ of the energy and momentum: in the relativistic case $\theta$ is the rapidity, $E(\theta):=m\cosh(\theta)$, $p(\theta):=m\sinh(\theta)$, and in the Galilean case $\theta$ is the velocity, $E(\theta):=m\theta^2/2$, $p(\theta):=m\theta$. A differential scattering phase $\varphi(\theta)$ fully specifies the dynamics of the model \cite{tba1}. We denote by $h_i(\theta)$ the one-particle eigenvalue of the conserved charge $Q_i$; in particular $h_1(\theta)=E(\theta)$ and $h_2(\theta)=p(\theta)$. 

Let us first recall the basic ingredients of TBA. Three related quantities play important roles: the quasi-particle density $\rho_{\rm p}(\theta)$, the state density $\rho_{\rm s}(\theta)$, and the quasi-particle occupation number $n(\theta):=\rho_{\rm p}(\theta)/\rho_{\rm s}(\theta)$. The functions $\rho_{\rm p}(\theta)$ and $n(\theta)$ are two different sets of state coordinates; each can be used to fully characterize the GGE. The former specifies all average densities in a simple way:
\beq\label{qrho}
	{\tt q}_i = \int \dd\theta\, \rho_{\rm p}(\theta) h_i(\theta).
\eeq
This can in fact be seen as a definition of $\rho_{\rm p}(\theta)$. Here and below, integrations are over $\R$.

As a consequence of interaction, quasi-particle and state densities are related to each other. Using the Bethe ansatz, one finds the following {\em constitutive relation} \cite{tba1}:
\beq
	2\pi\, \rho_{\rm s}(\theta) = p'(\theta) + \int \dd \alpha\,
	\varphi(\theta-\alpha)\rho_{\rm p}(\alpha)
\eeq
where $p'(\theta) = \dd p(\theta)/\dd \theta$. This relation gives rise to a nonlinear relation between the state coordinates $\rho_{\rm p}(\theta)$ and $n(\theta)$. The transformation from the former to the latter is direct from the above definitions. In the opposite direction, the transformation is effected by
\beq\label{rhon}
	2\pi\,\rho_{\rm p}(\theta) = n(\theta) \lt(p'\rt)^{\rm dr}(\theta)
\eeq
where the ``dressing'' operation $h\mapsto h^{\rm dr}$ is defined by the solution to the linear integral equation
\beq
	h^{\rm dr}(\theta) = h(\theta) + \int \frc{\dd\gamma}{2\pi}
	\varphi(\theta-\gamma)\,n(\gamma)\, h^{\rm dr}(\gamma).
	\label{q}
\eeq

The potentials $\underline\beta$ can be recovered: the occupation number is related to the one-particle eigenvalue $w(\theta)=\sum_i \beta_i h_i(\theta)$ of the charge $\sum_i \beta_i Q_i$ in the GGE \eqref{dens} via the so-called  pseudo-energy $\ep_w(\theta)$ \cite{tba1,moscaux}:
\beqa	\label{epth}
	n(\theta) &=& \frc1{1+\exp\big[\ep_{w}(\theta)\big]}\\
	\ep_w(\theta) &=& w(\theta) -
	\int \frc{\dd\gamma}{2\pi} \varphi(\theta-\gamma) \log(1+e^{-\ep_w(\gamma)}).\no
\eeqa

The above ingredients give exact average densities as functions of GGE states. However, they do not provide expressions for average currents as functions of state coordinates, and for equations of states. Hence they are not sufficient in order to develop generalized hydrodynamics.

We solve this problem by obtaining the following expressions:
\beq\label{qj}
	{\tt q}_i = \int \frc{\dd p(\theta)}{2\pi} n(\theta)
	h_i^{\rm dr}(\theta),\ \ 
	{\tt j}_i = \int \frc{\dd E(\theta)}{2\pi} n(\theta)
	h_i^{\rm dr}(\theta)
\eeq
where $h_i^{\rm dr}(\theta)$ is the dressed one-particle eigenvalue. These expressions emphasize the role of relativistic or Galilean symmetry: the sole difference between GGE averages of charge densities and currents is the integration measure, determined by the dispersion relation.

The first equation in \eqref{qj} is well known and is a consequence of \eqref{qrho} and \eqref{rhon}. In integral-operator notation (with measure $\dd\theta/(2\pi)$), the dressing operation is \beq\label{dressop}
	h^{\rm dr} = (1-\varphi {\cal N})^{-1}h
\eeq
where ${\cal N}$ is diagonal with kernel $2\pi\, n(\theta)\delta(\theta-\alpha)$, and $\varphi$ has kernel $\varphi(\theta-\alpha)$. Therefore, introducing the symmetric operator ${\cal U} = {\cal N}(1-\varphi {\cal N})^{-1}$ and the bilinear form $a\cdot b = \int \dd\theta/(2\pi)\, a(\theta)b(\theta)$, we have
\beq
	{\tt q}_i = h_i\cdot {\cal U}\, p' =
	p'\cdot{\cal U}\, h_i
\eeq
which leads to the first equation of \eqref{qj}.

The second equation in \eqref{qj} is new. It may be proven, in the relativistic case, using relativistic crossing symmetry, and then obtained by the non-relativistic limit in the Galilean case. In the relativistic case, crossing symmetry says that local currents $j_i$, in the cross-channel, are local densities $q_i$; therefore the current expression in \eqref{qj} is obtained from that of the density under an appropriate exchange of energy and momentum. Let ${\cal C}$ be the crossing transformation $(x,t)\mapsto (\ri t,-\ri x)$, implemented on rapidities by $\theta\mapsto \ri \pi/2-\theta$. Note that it squares to the identity, ${\cal C}^2 = 1$. Let us denote by $q[h]$ and $j[h]$ the density and current operators, respectively, associated to a one-particle eigenvalue $h(\theta)$. Then the statement that local currents $j_i$, in the cross-channel, are local densities $q_i$ translates into ${\cal C}(j[h]) = \ri q[h^C]$ where $h^C(\theta) = h(\ri\pi/2-\theta)$. Let us also denote by $\bra\Or\ket_w$ the average of observables $\Or$ in the state characterized by $w(\theta)$. Then $\bra{\cal C}(\Or)\ket_w = \bra\Or\ket_{w^C}$ where $w^C(\theta) = w(\ri\pi/2 - \theta)$. Using $\bra j[h]\ket_w = \bra {\cal C}({\cal C}(j[h]))\ket_w = \ri \bra q[h^c]\ket_{w^C}$ and the expression for ${\tt q}_i={\tt q}[h_i]$ in Equation \eqref{qj}, we obtain that for ${\tt j}_i = {\tt j}[h_i]$. An alternative proof, using the machinery of integrable systems, is presented in Appendix \ref{appcur}.

Expressions \eqref{qj} have interesting consequences. First, using ${\tt j}_i = h_i\cdot {\cal U}\, E'$ where $E'(\theta) = \dd E(\theta)/\dd\theta$ in \eqref{qj}, the average current may also be written in terms of a current spectral density $\rho_{\rm c}(\theta)$:
\beq\label{jrho}
	{\tt j}_i = \int \dd\theta \,\rho_{\rm c}(\theta)h_i(\theta),
\eeq
which takes the forms
\beq\label{rhoc}
	2\pi\,\rho_c(\theta) = n(\theta)\, (E')^{\rm dr}(\theta)= 2\pi\, v^{\rm eff}(\theta)\rho_{\rm p}(\theta).
\eeq
Here $v^{\rm eff}(\theta)$ is the {\em effective velocity}, defined by
\beq\label{vdr}
	v^{\rm eff}(\theta):=\frc{\lt(E'\rt)^{\rm dr}(\theta)}{\lt(p'\rt)^{\rm dr}(\theta)}.
\eeq
The effective velocity depends on the state via the occupation number entering the dressing operation, and brings out the quasi-particle interpretation of the current expression: since $\rho_c(\theta) = v^{\rm eff}(\theta)\rho_{\rm p}(\theta)$, quasi-particles are seen as moving at effective velocities $v^{\rm eff}(\theta)$, influenced by the state in which they move.

Second, one may extract explicit GGE equations of state from expressions \eqref{qj}. The equations of states are necessary and sufficient relations between densities and currents, guaranteeing the existence of $n(\theta)$ such that both relations in \eqref{qj} hold for all $h_i(\theta)$. Assume that ${\tt q}_i$ and ${\tt j}_i$ are averages in a state, not necessarily a GGE. In complete generality, both are linear functionals of $h(\theta)$, hence we may still write \eqref{qrho} and \eqref{jrho} for some quasi-particle density $\rho_{\rm p}(\theta)$ and current spectral density $\rho_{\rm c}(\theta)$. GGE equations of states can therefore be written as relations between $\rho_{\rm p}(\theta)$ and $\rho_{\rm c}(\theta)$, necessary and sufficient for the existence of $n(\theta)$ such that \eqref{qj} hold. One can show that these relations are
\beq
	\frc{\rho_{\rm c}(\theta)}{ \rho_{\rm p}(\theta)}
	= \frc{
	E'(\theta) + \int \dd \alpha\,
	\varphi(\theta-\alpha) \rho_{\rm c}(\alpha)
	}{
	p'(\theta) + \int \dd \alpha\,
	\varphi(\theta-\alpha) \rho_{\rm p}(\alpha)
	}.
	\label{tbaeos}
\eeq
These relations are independent of the state: they hold in any GGE, in the model described by the differential scattering phase $\varphi(\theta-\alpha)$. They characterize the set of doublets of functions $(\rho_{\rm p}, \rho_{\rm c})$ describing available GGEs for this integrable model. The proof of \eqref{tbaeos} is obtained by isolating $n(\theta)$ in both \eqref{rhon} and \eqref{rhoc}, in the forms $2\pi({\cal N}^{-1} - \varphi)\rho_{\rm p} = p'$ and $2\pi({\cal N}^{-1} - \varphi)\rho_{\rm c} = E'$, and equating the resulting expressions.

Finally, recalling \eqref{rhoc}, the left hand side of \eqref{tbaeos} is $v^{\rm eff}(\theta)$. Simple manipulations of \eqref{tbaeos} then give a linear integral equation for the effective velocity $v^{\rm eff}(\theta)$ in terms of quasi-particle densities:
\beq\label{veos}
	v^{\rm eff}(\theta) = v^{\rm gr}(\theta)
	+\int \dd\alpha\,\frc{\varphi(\theta-\alpha)\,
	\rho_{\rm p}(\alpha)}{
	p'(\theta)}\,
	(v^{\rm eff}(\alpha)-v^{\rm eff}(\theta))
\eeq
where $v^{\rm gr}(\theta) = E'(\theta)/p'(\theta)$ is the group velocity. In this form, the equations of state of integrable systems are seen as equations specifying an effective velocity of quasi-particles, as a modification of the group velocity.

We note that the effective velocity derived here agrees with that proposed in \cite{BEL14}. This is interesting, as our derivation is based on comparing current spectral density to quasi-particle density, while the concept proposed in \cite{BEL14} is based on stationary-phase arguments\footnote{We also note that \eqref{vdr} is {\em not} the derivative of the dressed energy with respect to the dressed momentum, as the $\theta$-derivative lies inside the dressing operation. It equals $\dd E^{\rm dr}(\theta)/ \dd p^{\rm dr}(\theta)$ if and only if the dressing is with respect to a pure state, and this is indeed the definition taken in \cite{BEL14}, where a representative pure state was chosen to describe the macrostate.}.

\subsection{Generalized hydrodynamics}

The basic equation of generalized pure hydrodynamics is derived from \eqref{conshy} along with the quasi-particle expressions \eqref{qrho} and \eqref{jrho}. The fact that the space of pseudolocal charges is complete \cite{Dtherm} suggests that these hold for a complete set of functions $h_i(\theta)$, and thus (here and below we suppress explicit $x,t$ dependences for lightness of notation):
\beq\label{gcons}
	\p_t \rho_{\rm p}(\theta) + \p_x \rho_{\rm c}(\theta) = 0.
\eeq
Using the equations of state \eqref{tbaeos}, this is an integro-differential system for the space-time dependent state characterized by the particle densities $\rho_{\rm p}(\theta)$.

Alternatively, using the dressed-velocity formulation \eqref{rhoc} and \eqref{veos}, Equation \eqref{gcons} may be written as
\beq\label{gcons2}
	\p_t \rho_{\rm p}(\theta) + \p_x \big(v^{\rm eff}(\theta)\rho_{\rm p}(\theta)\big) = 0.
\eeq
This is the conservation form of generalized hydrodynamics. It is a density-type conservation equation, and identifies $\rho_{\rm p}(\theta)$ as a conserved fluid density.

The state coordinates $\rho_{\rm p}(\theta)$ are, however, not the most convenient. We show that the occupation numbers $n(\theta)$ {\em diagonalize the Jacobian $J(\underline{\tt n})$} in the quasi-linear form \eqref{hydron}: the space-time dependent occupation number $n(\theta)$ satisfies the following integro-differential system, the vanishing of the convective derivative of $n(\theta)$:
\beq\label{ghydro}
	\p_t n(\theta) + v^{\rm eff}(\theta) \p_x n(\theta) = 0.
\eeq
Here \eqref{vdr} may be used to express the effective velocity in terms of $n(\theta)$. Hence, $n(\theta)$ are the normal modes of generalized hydrodynamics, and further, the eigenvalues -- the propagation velocities -- are exactly the effective velocities $v^{\rm eff}(\theta)$.

The proof of \eqref{ghydro} is as follows. Using the integral-operator relations $2\pi\rho_{\rm p} = {\cal U}p'$ and $2\pi\rho_{\rm c} = {\cal U}E'$, we have $(\p_t {\cal U})\,p' + (\p_x {\cal U})\,E' = 0$. Taking derivatives, $\p_{x,t} {\cal U} = (1-{\cal N} \varphi)^{-1}( \p_{x,t} {\cal N}) (1- \varphi {\cal N})^{-1}$, and we obtain
\beq\label{eig}
	\p_t {\cal N}\,(1-\varphi {\cal N})^{-1}p'
	+
	\p_x {\cal N}\,(1-\varphi {\cal N})^{-1}E'=0
\eeq
which gives \eqref{ghydro} using \eqref{dressop}.

Observe that using \eqref{gcons2} and \eqref{ghydro}, it is simple to show that the state density $\rho_{\rm s}(\theta)$, as well as the hole density $\rho_{\rm h}(\theta):=\rho_{\rm s}(\theta)-\rho_{\rm p}(\theta)$, also satisfy the {\em same density-type conservation equation} \eqref{gcons2}. Further, as a consequence, the {\em entropy density} \cite{tba1},
\beq
	s(\theta):= \rho_{\rm s}(\theta)\log \rho_{\rm s}(\theta) -
	\rho_{\rm p}(\theta) \log \rho_{\rm p}(\theta)
	- \rho_{\rm h}(\theta) \log \rho_{\rm h}(\theta),
\eeq
also satisfies this conservation equation, $\p_t s(\theta) + \p_x \big(v^{\rm eff}(\theta) s(\theta)\big)=0$. Conservation of entropy density is a fundamental property of perfect fluids, as no viscosity effects are taken into account.

In the large-scale limit the equation for the ray-dependent ($\xi$-dependent) occupation number $n(\theta)$ simplifies to
\[
	(v^{\rm eff}(\theta)-\xi)\,\p_\xi n(\theta)=0.
\]
This is the eigenvalue equation \eqref{hydroscale} in the occupation-number coordinates (which diagonalize the Jacobian), and its solution gives ${\underline {\tt q}}(\xi)$ and ${\underline {\tt j}}(\xi)$ via \eqref{qj}, \eqref{q}.

One can show that the solution for the non-equilibrium, ray-dependent occupation number $n(\theta)$ is the discontinuous function
\beq\label{n}
	n(\theta) = n^L(\theta) \Theta(\theta-\theta_\star)
	+n^R(\theta) \Theta(\theta_\star-\theta)
	\label{n}
\eeq
where $\Theta(\cdots)$ is Heavyside's step function. The position of the discontinuity $\theta_\star$ depends on $\xi$ and is self-consistently determined by $v^{\rm eff}(\theta_\star)=\xi$; equivalently, it is the zero of the dressed, boosted momentum $p_\xi(\theta) := p(\theta-\eta)$ where $\xi = \tanh\eta$ (relativistic case) or $\xi=\eta$ (Galilean case),
\beqa
	p_\xi^{\rm dr}(\theta_\star)=0.
	\label{y}
\eeqa
The GGE occupation numbers $n^{L,R}(\theta)$ entering \eqref{n} guarantee that the asymptotic conditions on $\xi$ correctly represent the asymptotic baths as per \eqref{hydroini}. They are obtained using \eqref{epth} with $w=w^{L,R}(\theta)$ the one-particle eigenvalues characterizing the GGE of the left and right asymptotic reservoirs; for instance, with reservoirs at temperatures $T_{L,R}$, we have $w^{L,R}(\theta) = T_{L,R}^{-1} \,E(\theta)$.

Indeed, the solution \eqref{n} of the scaled problem holds since $v^{\rm eff}(\theta)$ is monotonic and covers the full range of $\theta$  (which is $[-1,1]$ in the relativistic case and $\R$ in the Galilean case): therefore there is a unique solution to $v^{\rm eff}(\theta)=\xi$, thus a unique jump; and $\theta_\star$ is monotonic with $\xi$, hence the asymptotic conditions are correctly implemented.

The system of integral equations \eqref{qj}, \eqref{q}, \eqref{n} and \eqref{y} can be solved numerically using Mathematica, yielding extremely accurate results. Integral equations in \eqref{epth} and \eqref{q} can be solved iteratively, a procedure that converges fast \cite{tba1}. The hydrodynamic solution is obtained by first constructing the thermal occupation numbers $n^{L,R}(\theta)$ \eqref{epth}. Then, the non-equilibrium occupation number is evaluated by solving the system \eqref{n}, \eqref{y}: one first chooses $\theta_\star=\eta$ in order to construct $n(\theta)$, and then evaluates $p_\xi^{\rm dr}(\theta)$. The zero of $p_\xi^{\rm dr}(\theta)$ is numerically found -- we observed that $p_\xi^{\rm dr}(\theta)$ always has a single zero. The process is repeated until the zero is stable -- we observed that this is a convergent procedure. Finally, the non-equilibrium occupation number is used in \eqref{qj}, \eqref{q}. The solving time increases slowly with the numerical precision demanded, thus this allows arbitrary-precision results.

The solution presented may be interpreted as a single {\em space-covering rarefaction wave}, in the sense that it is a solution to the eigenvalue equation \eqref{hydroscale} where all physical observables ${\tt q}_i,\,{\tt j}_i$ are continuous and interpolate between the two reservoirs. With relativistic dispersion relation, the solution is smooth within the light cone, beyond which the states are constant and equal to the initial baths' states; while in the Galilean case, the solution is generically smooth on the whole space. In this solution, every normal mode $n(\theta)$, seen as a function of $\xi$ for fixed $\theta$, is discontinuous exactly at its propagation velocity. Every normal mode therefore displays a ``contact discontinuity'' (a discontinuity without entropy production) \cite{hydro}. Hence, the rarefaction wave may be seen as being composed of infinitely many contact discontinuities. In contrast to the finite-dimensional case, this single rarefaction wave can account for {\em generic reservoirs}, and no shock need to develop. This is  because in the infinite-dimensional case, the eigenvalues of $J(\underline{\tt n})$ form a continuum: all propagation velocities $v^{\rm eff}(\theta)$ are available as conserved charges guarantee a large number of stable excitations, providing an additional continuous parameter tuning the smooth state trajectory and guaranteeing its correct asymptotic-reservoir values.  Since weak solutions (shocks) are not necessary to connect the asymptotic reservoirs, they do not appear.

\section{Analysis and discussion}
Concentrating on pure thermal transport, we have analyzed the above general system of equations for two related models: the relativistic integrable sinh-Gordon model and its non-relativistic limit \cite{LLlimit1,LLlimit2}, the (repulsive) Lieb-Liniger model. We have also verified that our hydrodynamic equations reproduce the known results for the case of free particles.

\subsection{The relativistic sinh-Gordon model}
One of the simplest integrable relativistic QFT with non-trivial interactions is the sinh-Gordon model. 
It is defined by the Lagrangian
 \cite{toda2,toda1}:
\begin{equation}\label{lagran}
    \mathcal{L}_{{\rm shG}}=\frac{1}{2}(\partial_{\mu}\phi)^2-\frac{m^2}{\beta^2}\cosh(\beta\phi), \label{sg}
\end{equation}
where $\phi$ is the sinh-Gordon field and $m$ is the mass of the single particle in the spectrum.
 The model is integrable and therefore the only non-trivial scattering matrix is that associated to two-particle scattering. It is given by
\cite{SSG2,SSG3,SSG} 
\begin{equation}\label{smatrix}
    S(\theta)=\frac{\tanh\frac{1}{2}\left(\theta-\frac{\ri\pi B}{2}\right)}{\tanh\frac{1}{2}\left(\theta+\frac{\ri\pi B}{2}\right)}.
\end{equation}
The parameter $B \in [0,2]$ is the effective coupling constant
which is related to the coupling constant $\beta$ in the
Lagrangian by
\begin{equation}
\label{BB}
    B(\beta)=\frac{2\beta^2}{8\pi + \beta^2},
\end{equation}
under CFT normalization \cite{za}. The $S$-matrix is obviously
invariant under the transformation $B\rightarrow 2-B$, a symmetry
which is also referred to as weak-strong coupling duality, as it
corresponds to $B(\beta)\rightarrow B(\beta^{-1})$ in (\ref{BB}).
The point $B=1$ is known as the self-dual point. At the self-dual point the TBA differential scattering phase is simply
\beq 
\varphi_{{\rm shG}}(\theta)=-\ri\frc{\dd}{\dd\theta} \log S(\theta) = \frac{2}{\cosh\theta}.
\eeq

Contrary to the Lieb-Liniger model which we will discuss later, the general features of any quantities of interest in the sinh-Gordon model are very similar for any values of the coupling $B$. For this reason,
 in this paper we will concentrate our analysis solely on the self-dual point in the understanding that similar results hold for other values of $B$.

\begin{figure}[h!]
\begin{center}
\includegraphics[width=8cm]{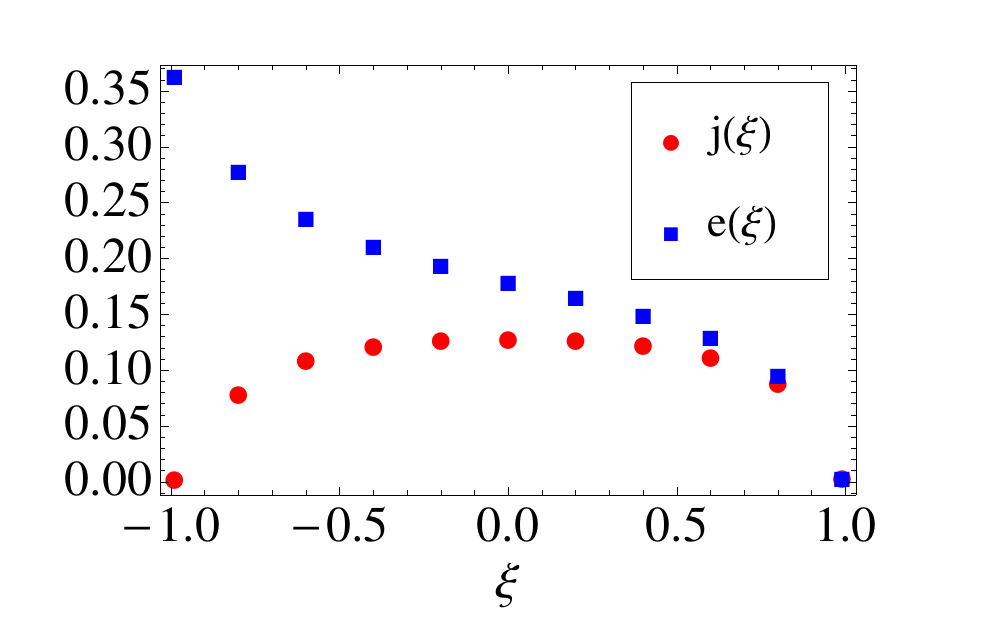}
\caption{The functions ${\tt j}(\xi)$ (dots) and ${\tt e}(\xi)$ (squares) for $\beta_L=1$ and $\beta_R=30$ in the sinh-Gordon model. }
\label{figcur1}
\end{center}
\end{figure}

\begin{figure}[h!]
\begin{center}
\includegraphics[width=7.5cm]{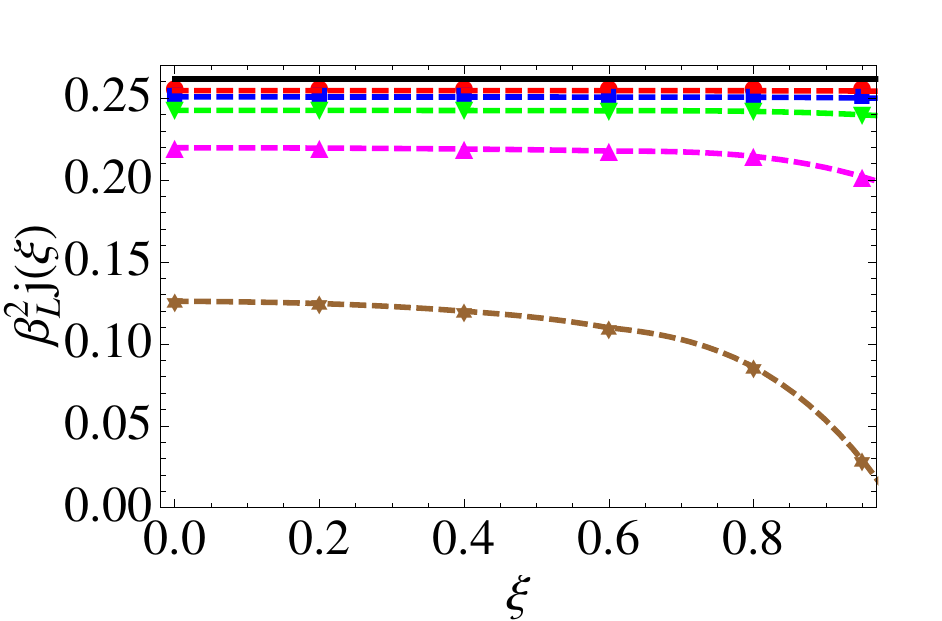}
\caption{The functions $\beta_L^2{\tt j}(\xi>0)$ for $\beta_R=30\beta_L$ and $\beta_L=10^{-p}$ with $p=0$ (stars), $1$ (standing triangles), $2$ (inverted triangles), $3$ (squares) and $4$ (circles). The continuous bold line represents the conformal value $\beta_L^2 {\tt j}(\xi)=\frac{\pi}{12}\left(1-\frac{1}{900}\right)$ which, as expected is reached at high temperatures. Dashed curves are interpolations.}
\label{figcur2}
\end{center}
\end{figure}

\begin{figure}
\includegraphics[width=7.5cm]{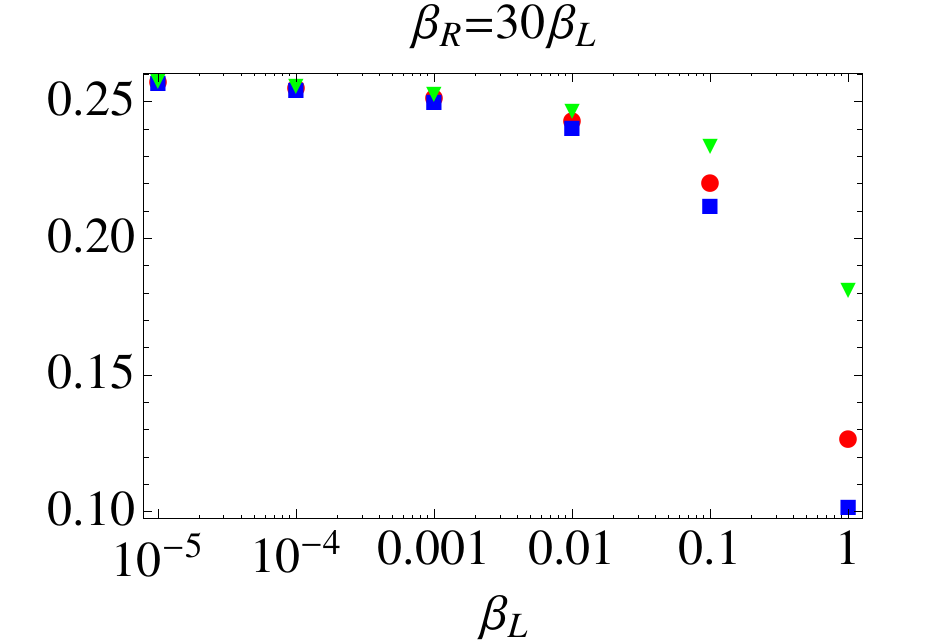}
\caption{Verification of the inequalities \eqref{ineq} in the sinh-Gordon model. Displayed are the functions $\beta_L^2 {\tt j}^{\rm sta}$ (circles),  $\beta_L^2({\tt e}^L-{\tt e}^R)/2$  (triangles) and $\beta_L^2 ({\tt k}^L-{\tt k}^R)/2$ (squares).}
\label{figdev}
\end{figure}

We have evaluated the energy density ${\tt e}:={\tt q}_1$, energy current ${\tt j}:={\tt j}_1$ and pressure ${\tt k}:={\tt j}_2$. Typical profiles are shown in Figs. \ref{figcur1}, \ref{figcur2}. Fig.~\ref{figcur1} shows smooth interpolation within the light cone between the asymptotic baths at $\xi=-1$ and $\xi=1$ (the speed of light is set to 1). Fig.~\ref{figcur2} shows how, as temperatures rise, the current approaches the plateau \eqref{CFT} predicted by CFT \cite{BD2012,BD-long}. Further, in Fig. \ref{figdev}, the relative deviation of the steady-state current from its bounds \eqref{ineq} is shown. The bounds are extremely tight, pointing to the strength of this constraint and confirming that the proposed solution is correct.
The numerical data have been obtained by solving the integral equations recursively until convergence is reached.
Sources of error are the discretization and finite range of $\theta$ for numerical integration. Adjusting the number of divisions and the range, we estimate the error to be less than $0.1\%$.

\subsection{The Lieb-Liniger model}
The Lieb-Liniger (LL) model, in the repulsive regime ($\lambda>0$), can be regarded as a non-relativistic limit of the sinh-Gordon model, as shown in \cite{LLlimit1,LLlimit2}. 
The Hamiltonian of the model is given by
\begin{equation}
 H_{\mathrm{LL}}=\int \dd x
 \Bigl(\frac{1}{2m}\p_x\psi^{\dagger}\p_x\psi+\lambda\psi^{\dagger}\psi^{\dagger}\psi\psi\Bigr).
\end{equation}
This is obtained from the Hamiltonian of the sinh-Gordon model by a double limit
 \begin{equation}
 c \to \infty,\ \beta \to 0;\ \beta c=4\sqrt{\lambda},
\end{equation}
where $c$ is the speed of light (which was implicit in
(\ref{sg}))\footnote{This is the only equation in the present paper where the speed of light $c$ appears explicitly. Everywhere else $c$ denotes the central charge. As both are
standard notations, we opted to use the name $c$ for both.}.  This limit can be performed within the TBA formalism \cite{LLlimit2}, and accordingly, the density  and
current averages $\qq_{i}$, $\jj_{i}$ are given by \eqref{qj}, with the non-relativistic dispersion relation. There, the occupation number is given by
$n_{\LL}(\theta)=1/(1+e^{\epsilon_{w}(\theta)})$, and the pseudo-energy
 $\epsilon_{w}(\theta)$ and the dressed one-particle eigenvalues
 $h^{\dr}_{i}(\theta)$ are defined in the same manner as in Equations \eqref{epth} and \eqref{q} (where $\theta=p/m$ is the velocity), with scattering matrix given by
\beq 
S_{LL}(\theta)=\frac{\theta-2\lambda i}{\theta+ 2\lambda i},
\label{ss}
\eeq 
and corresponding differential scattering phase
\begin{equation}
 \varphi_{\LL}(\theta)=\frac{4\lambda}{\theta^2+4\lambda^2}.
 \label{kernel}
\end{equation}
A uniform chemical potential $\mu$ is introduced, associated to the conserved charge $Q_0$ that counts the number of quasi-particles (with $h_0(\theta)=1$). The energy current is chosen to be the current associated to of charge $H-\mu Q_0$,
\beq
	{\tt j} := {\tt j}_1 - \mu {\tt j}_0\qquad\mbox{(LL model)}.
\eeq
Below we present some numerical results for several values of the coupling $\lambda$ and for $m=1$.

\begin{figure}
\includegraphics[width=7.5cm]{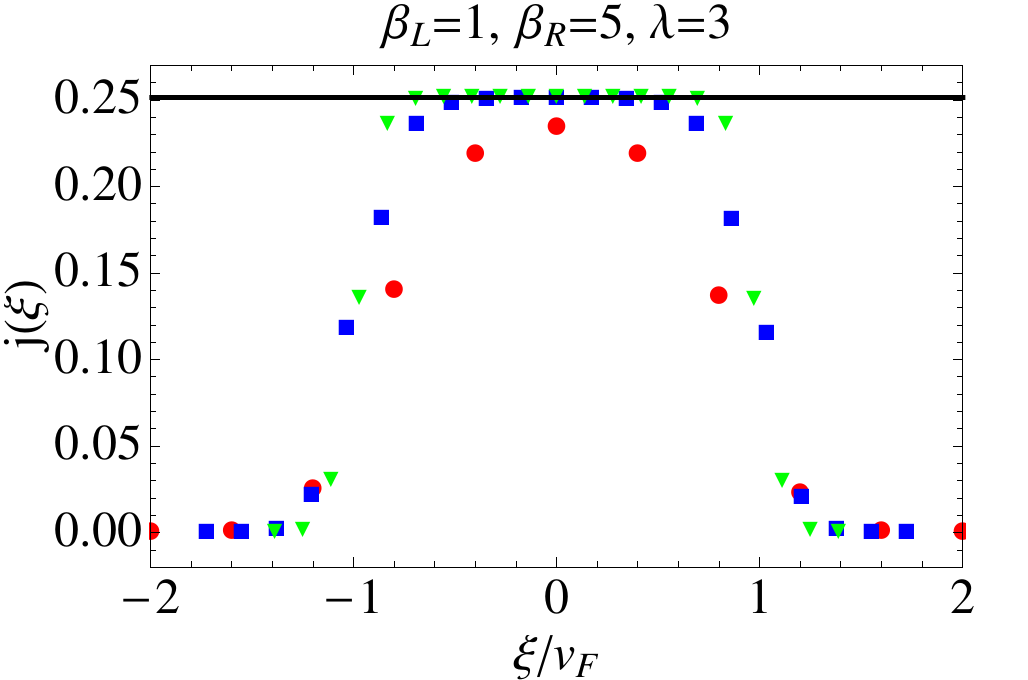}
\caption{Energy current in the Lieb-Liniger model for low temperatures, $\lambda=3$ and chemical potentials $\mu=3$ (circles), $\mu=6$ (squares) and $\mu=10$ (triangles). 
The CFT value $\frac{\pi}{12}\left(1-\frac{1}{25}\right)$ (bold horizontal line) is reached for high values of $\mu$. By plotting the currents against $\xi/v_F$ we observe the collapse of the various curves, which becomes better as $\mu$ increases. 
The regions where plateaux emerge are roughly $\xi/v_F \in [-1, 1]$ with $v_F\approx 2.5, 2.9, 3.6$. }
\label{figLL}
\end{figure}

Current profiles obtained for $\lambda=3$ and various values of $\mu$ are displayed in Fig.~\ref{figLL}. The main difference between the relativistic and non-relativistic cases is the lack, in the latter, of any sharp light-cone effect. Nevertheless, at low temperatures $T_{L,R}\ll\mu$, Luttinger Liquid physics emerges \cite{lowTLL}, including an emerging light-cone due to the Fermi velocity. This can be seen in Fig.~\ref{figLL}: a plateau forms whose height is again in agreement with the general CFT result \eqref{CFT}. The plateau lies between nearly symmetric values $\xi/v_F\approx \pm 1$ fixed by the Fermi velocity $v_F$. Thermal occupation numbers $n^{L,R}(\theta)$ are very sharply supported between Fermi points $\pm\theta^{L,R}_F$ with $\theta^{L,R}\gtrsim \sqrt{2\mu/m}$, and the Fermi velocity, which depends on $\xi$ very weakly, is the effective velocity $v^{\rm eff}(\theta_F^R)$ associated to the lowest temperature ($T_R<T_L$ in the present example). In agreement with general CFT results \cite{BD2012,BD-long}, a light cone thus builds up (despite the model having no intrinsic maximal velocity), and the full state is in fact homogeneous between the Fermi velocities.

In the LL model the coupling constant may take any values between 0 and $\infty$ and the limits $\lambda \rightarrow 0$ and $\lambda \rightarrow \infty$ are of particular interest.

\begin{figure}
\includegraphics[width=7cm]{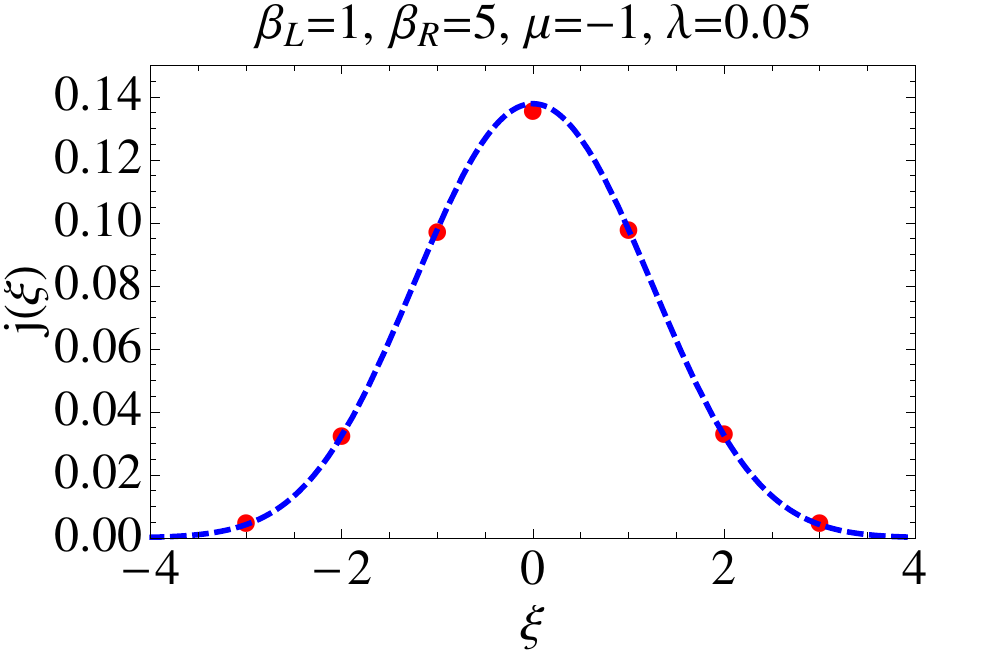}
\caption{Energy current in the Lieb-Liniger model for low temperatures, small coupling and negative chemical potential  (circles). 
The dashed curve represents the current (\ref{fb}) for the same temperatures and chemical potential. }
\label{smallL}
\end{figure}

For $\lambda\rightarrow 0$ the differential scattering phase  (\ref{kernel}) becomes heavily peaked around $\theta=0$. Formally,  $\lim_{\lambda \rightarrow 0}\varphi_{LL}(\theta)=2\pi \delta(\theta)$. The resulting TBA equations, with this differential scattering phase, admit no solution for the pseudoenergy if $\mu>0$, but for $\mu<0$ they can be solved exactly and reproduce the free Boson solution (for which $\mu>0$ would make no physical sense). In particular the energy current takes the free Boson form,
\begin{eqnarray} 
\!\!\!\! \!\lim_{\lambda \rightarrow 0} {\tt j}(\xi) =\frac{1}{\beta_R^2}\int_{\alpha_R}^\infty 
 d\theta \, \frac{\theta}{e^{\theta}-1}-\frac{1}{\beta_L^2}\int^{\infty}_{\alpha_L}  d\theta \, \frac{\theta}{e^{\theta}-1}, \label{fb}
\end{eqnarray} 
where $\alpha_{L,R}=\beta_{L,R}(\frac{\xi^2}{2}-\mu)$. In Fig.~\ref{smallL} we compare numerical values for $\lambda=0.05$ and
$\mu=-1$ to this analytical expression. The agreement is very good,
confirming that a free Boson theory is smoothly recovered in this limit.
With $\mu>0$, as $\lambda$ becomes small the TBA equations gradually breakdown. How this occurs is subtle, and will be discussed in \cite{CDYfuture}.

The qualitative change in behaviour of the TBA solutions as
$\lambda\to0$ is related to the two distinct regimes observed at small
values of $\lambda$ \cite{GP}. Consider the dimensionless coupling
$\gamma := 2m\lambda/{\tt q}_0$ (where we recall that ${\tt q}_0$ is the
gas density, which may be taken in the initial baths for instance) and the reduced temperature $\tau := 2 mT/{\tt q}_0^2$. The
``decoherent regime'', with large phase and density fluctuations, occurs
for $\gamma \lesssim {\rm min}(\tau^2,\sqrt{\tau})$. In this regime,
ideal Bose gas physics is recovered, and we have indeed verified that
the inequality is satisfied in the parameter space where good agreement
with \eqref{fb} is observed (small $\lambda$, negative $\mu$). On the
other hand, the ``Gross-Pitaevskii regime'' occurs for
$\tau^2\lesssim\gamma \lesssim1$, a quasi-condensate with large phase fluctuations but suppressed density fluctuations. It is such quasi-condensate physics that strongly affects TBA solutions as $\lambda\to0$ with $\mu>0$.

\begin{figure}
\includegraphics[width=7.5cm]{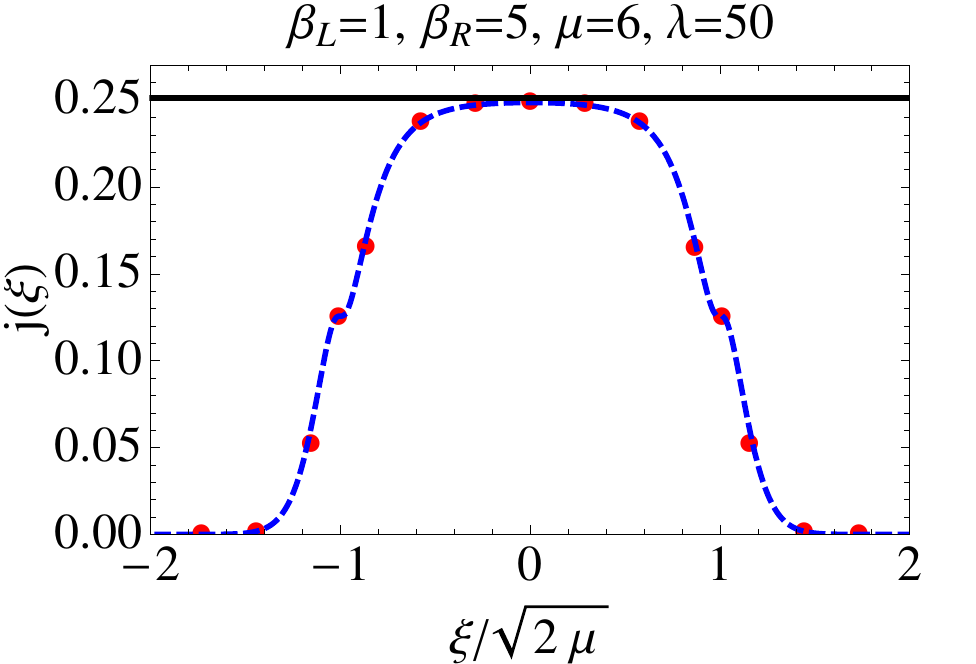}
\caption{Energy current in the Lieb-Liniger model for low temperatures, large coupling and chemical potential $\mu=6$ (circles). Local stationary points occur at $\alpha_{L,R}=0$, that is $\xi=\pm \sqrt{2\mu}=\pm 3.46$ (the Fermi velocity).
The dashed curve represents the current (\ref{ff}) for the same temperatures and chemical potential, whose profile is not dissimilar to the plots shown in Fig.~\ref{figLL}. As before, the bold horizontal line is the CFT value $\frac{\pi}{12}\left(1-\frac{1}{25}\right)$. 
The agreement is extremely good. }
\label{largeL}
\end{figure}

The other interesting limit is $\lim_{\lambda \rightarrow \infty} \varphi_{LL}(\theta)=0$. In this case we can also find an analytical expression for the current:
\begin{eqnarray} 
\!\!\!\! \!\!\lim_{\lambda \rightarrow \infty} {\tt j}(\xi) =\frac{1}{\beta_R^2}\int_{\alpha_R}^\infty 
 d\theta \, \frac{\theta}{e^{\theta}+1}-\frac{1}{\beta_L^2}\int^{\infty}_{\alpha_L}  d\theta \, \frac{\theta}{e^{\theta}+1}. \label{ff}
\end{eqnarray} 
This corresponds to a free Fermion, in agreement with the expected Tonks-Girardeau physics occuring in the regime $\gamma \gtrsim {\rm max}(1,\sqrt{\tau})$ \cite{GP}. For $\xi \approx 0$ and $\mu \beta_{L,R}\gg 1$  it is easy to show that the integral above gives $\frac{\pi}{12}(\beta_L^{-2}-\beta_R^{-2})$ so that we recover the CFT result for the current with $c=1$ (Dirac Fermion).
Fig.~\ref{largeL} shows a comparison between numerical values of the current for $\lambda=50$ and the formula above.

\begin{figure}
\includegraphics[width=7.5cm]{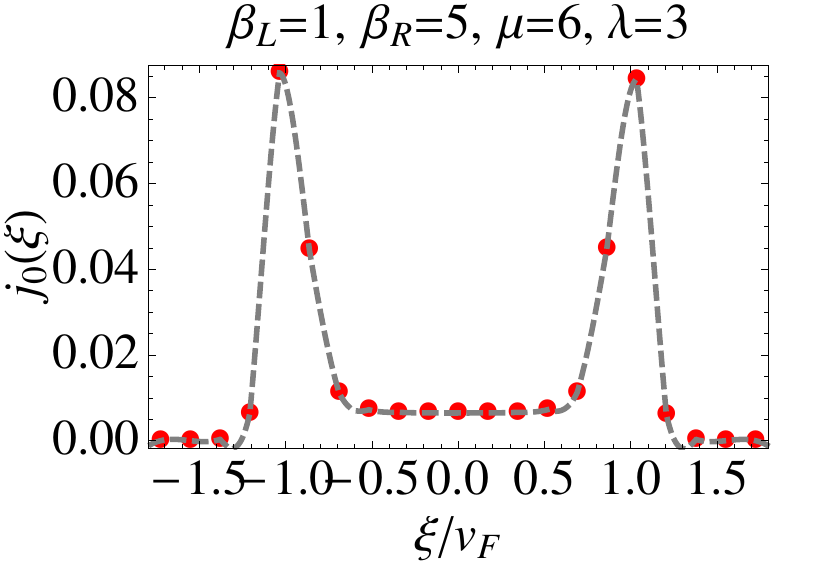}
\caption{A characteristic profile of the Lieb-Liniger particle current for $T_{L,R}\ll\mu$, $\lambda=3$ and $\mu=6$. The local maxima/minima are located around $\xi=\pm v_F$. The dashed curve is an interpolation.}
\label{j0mu6}
\end{figure}

Let us now consider the particle current. Naturally, in the LL model, equilibrium states at higher temperatures have lower particle densities. Therefore, although the energy current flows from the left to the right in the present setup (with $T_L>T_R$), the initial particle density imbalance would naively suggest a particle flow from the right (higher density) to the left (lower density). The {\em opposite} occurs: Fig.~\ref{j0mu6} shows that the particle current is positive, hence flows form the left to the right. This means that the fluid flow produced by the temperature difference drags particles with enough force to counteract the particle imbalance and bring particles towards the higher-density bath. The fact that heat carries particles along its motion is a {\em thermoelectric effect}. It has been experimentally demonstrated in a quasi-two-dimensional fermionic cold atoms channel \cite{Brant13}, and theoretically shown in CFT in dimensions higher than one \cite{rarefact1}. It is nontrivial in integrable models, as the large amount of conservation laws allows for independent currents to coexist, and our result gives the first theoretical prediction of this effect in the integrable one-dimensional Bose gas.

\begin{figure}
\includegraphics[width=7.5cm]{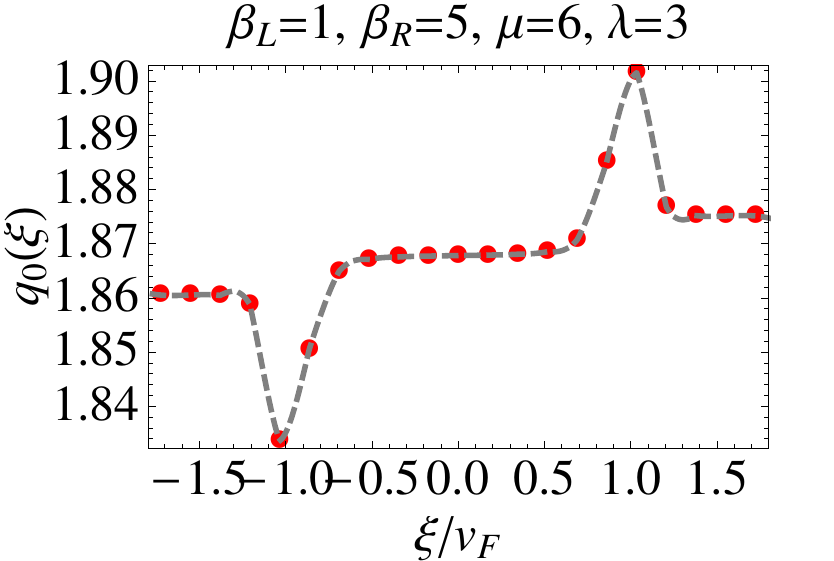}
\caption{A characteristic profile of the Lieb-Liniger particle density for $T_{L,R}\ll\mu$, $\lambda=3$ and $\mu=6$. The local maxima/minima are located around $\xi=\pm v_F$. The dashed curve is an interpolation.}
\label{q0mu6}
\end{figure}

An additional consequence of the thermoelectric effect is that the particle density ${\tt q_0}(\xi)$ shows particle accumulation around
$v_F$ and depletion around $-v_F$ (see Fig. \ref{q0mu6}). For instance,
the start of the dip can be explained by the fact that, in
any local spacial region originally in the left reservoir, the first particles
to start moving towards the right are those on the right of the region,
escaping and thus depleting it. Since time evolution at fixed position is obtained by scanning Fig.~\ref{q0mu6} from left to right, this explains the initial dip on the left. This depleting effect continues as long as the outgoing current on the right of the region is higher then the incoming current on its left -- that is, until the region lies in the steady state. However, as time goes on, the effective local temperature decreases, and this tends to increase the particle density. This effect eventually overtakes the depleting effect, accounting for the rebounce to the higher steady-state value. The behavior of the current ${\tt j}_0$ in Fig. \ref{j0mu6} is then a consequence of the continuity equation $\xi\p_{\xi}{\tt
 q}_0=\p_{\xi}{\tt j}_0$.

This is a nonuniversal effect, not present in the density ${\tt q}_1(\xi)-\mu {\tt q}_0(\xi)$ controlled by low-energy processes, where the physics of chiral separation dominates and monotonic transition regions occur.

\subsection{General features}
\begin{figure}
\includegraphics[width=7.5cm]{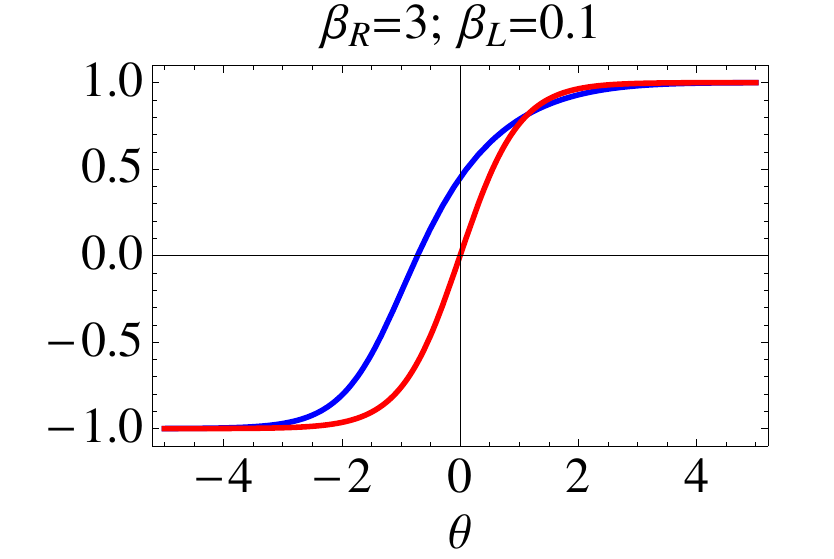}
\caption{Effective velocity in the sinh-Gordon model for $\xi=0$. Displayed are the effective velocity $v^{\rm eff}(\theta)$ (blue line) and the bare relativistic velocity $\tanh\theta$ (red line).}
\label{groupvel}
\end{figure}

The form of the non-equilibrium occupation number indicates that quasi-particles are thermalized according to the initial GGE's, in a way that depends on the rapidity. It connects with the picture, proposed in \cite{BD2012,castroint}, according to which in the steady state ($\xi=0$), quasi-particles traveling towards the right (left) are thermalized according to the left (right) reservoir. However, in the present solution, what determines the traveling direction is the {\em effective velocity} in the steady state: quasi-particles with positive (negative) dressed velocities, reaching the point $x=0$ at large times, will have travelled mostly towards the right (left) (after a relatively small transient). In the sinh-Gordon model with $T_L> T_R$, the effective velocity behaves as in Fig.~\ref{groupvel}. We observe that it is greater than the bare velocity $\tanh\theta$ for small or negative rapidities, and smaller for large positive rapidities. This is in agreement with the intuition according to which the quasi-particles are effectively carried by the flow, which transports them towards the right, for small enough rapidities, but slowed down by dominant ``friction'' effects of thermal fluctuations at large rapidities. A similar effects occur in the Lieb-Liniger model.

The generalized hydrodynamic result differs from previous proposals in interacting integrable models \cite{castroint,DeLucaVitiXXZ,Zotos16} (while all results agree in noninteracting cases). The original proposal \cite{castroint} was later shown \cite{D_2014} to break the second inequality in \eqref{ineq}, while the second proposal \cite{DeLucaVitiXXZ}, based on similar ideas, gave slight disagreements with numerical simulations. The conjecture \cite{Zotos16} which corresponds to taking $\theta_\star=0$ in our framework, seems to give good agreement with numerical simulations. This may be due to the fact that taking $\theta_\star=0$, gives very small errors in wide temperature ranges, of the order of 0.5-1\% (we have confirmed this numerically).

\section{Conclusions}

In this paper we developed a hydrodynamic theory for infinitely-many conservation laws, and applied it to the study of heat flows in experimentally relevant integrable models. It would be interesting to study further the non-equilibrium physics of the Lieb-Liniger model, including the effects of the Gross-Pitaevskii quasi-condensate and transport between different regimes. The emerging physical picture and solution we have given can be applied to any Bethe-ansatz integrable model, where infinitely-many conservation laws exist and a quasi-particle description is available. This includes quantum chains (see \cite{BCDF16}), as continuity of space on which the microscopic theory lies is not needed for emerging hydrodynamics. It also includes relativistic models with non-diagonal scattering such as the sine-Gordon model, where, for instance, our TBA construction may be generalized along the lines of the famous approach of Destri and de Vega \cite{destri1,destri2}. Of course, the hydrodynamic ideas do not require a quasi-particle description, and it might be possible to develop generalized hydrodynamics using a variety of techniques from integrability. We note that it is remarkable that independent quasi-particle mode thermalization agrees, in integrable models, with local entropy maximization. The dynamical equations derived can be used to describe more general situations in ultracold gases such as the release from a trap (see e.g. \cite{expgas}). This new theory and its extensions, including viscosity effects and forcing, should also allow for efficient studies of integrability breaking and related problems in any dimensionality, as well as for exact descriptions of dynamics in smooth trapping potentials \cite{KWW} at arbitrary coupling strength.

\vspace{0.1cm}
\noindent {\bf Acknowledgments.}\\ \noindent We thank Denis Bernard, M. Joe Bhaseen, Fabian Essler, Mauricio Fagotti and Eric Lutz for discussions, and especially Bruno Bertini and Lorenzo Piroli for sharing their preliminary ideas and Pasquale Calabrese for encouraging us to pursue this research. OAC-A and BD are grateful to SISSA, Trieste, Italy, for support during a visit where this work started. TY thanks the Takenaka Scholarship Foundation for a scholarship.

\appendix

\section{Current generators}\label{curpot}

Let $\bra\cdots\ket_{\underline\beta}$ be the state given by Equation \eqref{dens}, and $\bra a(x)b(y)\ket^{\rm c}:=  \bra a(x)b(y)\ket_{\underline\beta} -  \bra a(x)\ket_{\underline\beta} \, \bra b(y)\ket_{\underline\beta}$ the connected correlation functions. These are time-independent and functions of the difference $x-y$ only. Let us assume that connected correlation functions of conserved densities and currents vanish {\em faster} than the inverse distance $|x-y|$. Then,
\beqa
	\int \dd x\,\bra q_m(x) j_n(0)\ket^{\rm c}
	&=& \int \dd x\,\bra j_n(0) q_m(x)\ket^{\rm c} \n
	&=& \int \dd x\,\bra j_n(x) q_m(0)\ket^{\rm c} \n
	&=& -\int \dd x\,x\,\bra \p_xj_n(x)q_m(0)\ket^{\rm c} \n
	&=& \int \dd x\,x\,\bra \p_tq_n(x)q_m(0)\ket^{\rm c} \n	
	&=& -\int \dd x\,x\,\bra q_n(x)\p_tq_m(0)\ket^{\rm c} \n	
	&=& \int \dd x\,x\,\bra q_n(x)\p_yj_m(y)\ket^{\rm c}|_{y=0} \n	
	&=& -\int \dd x\,x\,\p_x \bra q_n(x)j_m(0)\ket^{\rm c} \n	
	&=& \int \dd x\,\bra q_n(x)j_m(0)\ket^{\rm c}.
\eeqa
In the first line we used the fact that $\int \dd x\,q_m(x)$ is a conserved quantity and thus commutes with the density matrix $\rho$; in the second we used space translation invariance, in the third integration by parts and the fast-enough vanishing of correlation functions; in the fourth current conservation; in the fifth time-translation invariance; in the sixth current conservation, in the seventh space-translation invariance; and in the eight integration by parts. Therefore,
\beq
	\frc{\p}{\p \beta_m} {\tt j}_n = \frc{\p}{\p \beta_n} {\tt j}_m
\eeq
and thus
\beq
	{\tt j}_m = \frc{\p}{\p \beta_m} g_{\underline\beta}
\eeq
showing Equation \eqref{fe}.

In the TBA context, we note that expressions \eqref{qj} show the existence of appropriate free energies $f_w$ and $g_w$ generating densities and currents, respectively, 
as in \eqref{fe}. Indeed they may be re-written as
\beq
	{\tt q}_i = \int \dd\theta\, h_i(\theta)
	\frc{\delta f_w}{\delta w(\theta)},\ 
	f_w = -\int \frc{\dd p(\alpha)}{2\pi} 
	\log(1+e^{-\ep_w(\alpha)}) \label{fw}
\eeq
and
\beq
	{\tt j}_i = \int \dd\theta\, h_i(\theta)
	\frc{\delta g_w}{\delta w(\theta)},\ 
	g_w = -\int \frc{\dd E(\alpha)}{2\pi}
	\log(1+e^{-\ep_w(\alpha)}).\label{gw}
\eeq
It then follows that that functional $w(\theta)$-derivatives of these free energies give the quasi-particle and current densities, $\rho_{\rm p}(\theta) = \delta f_w / \delta w(\theta)$ and $\rho_{\rm c}(\theta) =  \delta g_w / \delta w(\theta)$.

\section{Emergence of generalized hydrodynamics}\label{apparg}

The only principle at the basis of hydrodynamics, and of the derivation we provide, is that of the emergence of local generalized thermalization (local entropy maximization). Technically, this is the assumption that averages of local quantities $\bra \Or(x,t)\ket$ tend uniformly enough, at large $x$ and $t$, to averages evaluated in GGEs (infinite-volume, maximal-entropy states, under conditions on infinitely-many conservation laws), with space-time-dependent potentials. This assumption is sufficient to derive the explicit dynamics for all single-point averages of local conserved densities and currents: no ad-hoc kinetic principle is needed.

In the case of infinitely-many conservation laws, one delicate point is the consideration of quasi-local densities and currents, which are involved in generalized thermalization. Such a quantity is not supported on a finite region, but on an extended region, with a weight (as measured by, for instance, the overlap with any other local observable) that decays away from a point. However, since hydrodynamics is concerned with large-scale space-time regions (the fluid cells), it is natural to consider them on the same footing. This is implicitly done in the derivation presented in this paper by assuming a completeness property of conservation laws.

Another delicate point concerns the definition of GGEs itself. In finite systems, such states depend on the boundary conditions imposed, and in general, these boundary conditions may still have an effect in the infinite-volume limit. For instance, walls simply preclude any nonzero potential associated to the momentum operator, as they break translation invariance. Nevertheless, given a set of allowed conserved charges, at large volumes, boundary conditions have little effect on local averages of conserved currents and densities (as they do not affect specific free energies). Further, periodic boundary conditions, at the basis of the TBA formalism, appear to provide the maximal set of conserved charges. It is in fact possible to construct GGEs directly in infinite volumes \cite{Dtherm}. We expect local thermalization, and the full set of available conserved charges, to be correctly described by such constructions; and we expect these to agree with the TBA formalism used here.

We finally mention that the classical hard-rod problem, proven to give rise to a form of hydrodynamics \cite{dobrods,spobook}, has strong connections with the integrable systems investigated here, which we will investigate in a future work.

\section{Many-particle spectrum}\label{morepart}

The theory developed here is directly applicable to any integrable model whose two-particle scattering is diagonal in the internal space. Let the spectrum be composed of $\ell$ particles, of masses $m_a,\,a=1,\ldots,\ell$, and assume that their scattering is diagonal. In this case, the TBA equations can still be applied \cite{tba1,tba2}: the differential scattering phase is replaced by a matrix of functions $\varphi_{ab}(\theta-\gamma)$, and the one-particle eigenvalue of $Q_i$ will be denoted by $h_{i}(\theta;a)$. The solution $\underline{{\tt q}}(\xi)$, $\underline{\tt j}(\xi)$ of the generalized hydrodynamic problem is:
\beqa
	{\tt q}_i(\xi) &=& \sum_a \int \frc{\dd p(\theta;a)}{2\pi} n(\theta;a)
	h_{i}^{\rm dr}(\theta;a)\n
	{\tt j}_i(\xi) &=& \sum_a \int \frc{\dd E(\theta;a)}{2\pi} n(\theta;a)
	h_{i}^{\rm dr}(\theta;a)
\eeqa
where $p(\theta;a) = m_a\sinh\theta$ and $E(\theta;a) = m_a\cosh\theta$, and
\beq
	h_{i}^{\rm dr}(\theta;a) = h_{i}(\theta;a) + \int \frc{\dd\gamma}{2\pi}
	\sum_b \varphi_{a,b}(\theta-\gamma)\,n(\gamma;b)\, h_{i}^{\rm dr}(\gamma;b).
\eeq
The non-equilibrium occupation number $n(\theta;a)$ is given by the discontinuous function
\beq\label{napp}
	n(\theta;a) = n^L(\theta;a) \Theta(\theta-\theta_\star(a))
	+n^R(\theta;a) \Theta(\theta_\star(a)-\theta)
\eeq
where each particle $a$ is associated to a different discontinuity at position $\theta_\star(a)$. These positions are self-consistently determined by the zeroes of the dressed, boosted momenta of particles $a$; with $p_\xi(\theta;a) := m_a\sinh(\theta-\eta)$ (relativistic) or $m_a\theta$ (non-relativistic):
\beqa
	p_\xi^{\rm dr}(\theta_\star(a);a)=0,\quad a=1,\ldots,\ell.
\eeqa
Again the thermal occupation numbers $n^{L,R}(\theta;a)$ entering \eqref{napp} guarantee that the asymptotic conditions on $\xi$ correctly represent the asymptotic baths as per Equation \eqref{hydroini}. They are obtained using the TBA equations in terms of the pseudo-energies $\ep_w(\theta;a)$ \cite{tba1,tba2},
\beqa
	n^{L,R}(\theta;a) &=& \frc1{1+\exp\big[
	\ep_{w^{L,R}}(\theta;a)\big]}\\
	\ep_w(\theta;a) &=& w(\theta;a) - \n &&
	-\int \frc{\dd\gamma}{2\pi} \sum_b\varphi_{a,b}(\theta-\gamma) \log(1+e^{-\ep_w(\gamma;b)}).\no
\eeqa
Here $w^{L,R}(\theta;a)=\sum_i \beta_i^{L,R}h_i(\theta;a)$ are the one-particle eigenvalues of the charge $\sum_i \beta_i^{L,R}Q_i$ characterizing the GGE of the left and right asymptotic reservoirs.

\section{Current averages}\label{appcur}

An alternative proof  of Equations (\ref{qj}) may be given using the technology of integrable systems, which has the advantage of generalizing to flows generated by any conserved charge instead of just  the Hamiltonian. For completeness we present here the main arguments. The idea is to prove expression \eqref{qj} for current averages ${\tt j}_i$ given the expression for density averages ${\tt q}_i$. This is akin to extending the LeClair-Mussardo
formula (LM formula) \cite{LM} so that it incorporates the infinite number of
 conserved charges, and applying it to the current with the aid of form
factors (FFs). We shall use the notation introduced in \cite{Mussardo}. Following the derivation in \cite{Mussardo} we generalize the LM formula for a one-point function of
a generic local operator
$\mathcal{O}(x,t)$,
\begin{equation}
 \langle \mathcal{O}(x,t)\rangle = \sum_{\ell=0}^{\infty}\frac{1}{\ell!}\biggl(\prod_{k=1}^{\ell}\frac{\dd
  \theta_k}{2\pi}n(\theta_k)\biggr)\langle
  \ltheta|\mathcal{O}(0)|\rtheta \rangle_{\mathrm{c}},
\end{equation}
where $|\rtheta \rangle = |\theta_1,\cdots,\theta_\ell \rangle$ (and $\langle \ltheta |=\langle \theta_\ell,\cdots,\theta_1|$ is its hermitian conjugate) and diagonal matrix elements (DMEs) in the sum are connected (the meaning of being
 ``connected'' will be described below). Here $n(\theta)$
 is the same occupation number as that involved in \eqref{epth}, \eqref{qj}, \eqref{q}. It is then immediate to
 see that an expression for the density average ${\tt q}_i$ with the one-particle eigenvalue $h_i(\theta)$
 proved by Saleur \cite{Saleur} is modified to
\beqa
{\tt q}_i&=&m\sum_{\ell=0}^{\infty}\biggl(\prod_{k=1}^{\ell}\frac{\dd
  \theta_k}{2\pi}n(\theta_k)\biggr) \nonumber \\
         &\times& \varphi(\theta_{12})\cdots
  \varphi(\theta_{\ell-1,\ell})h_i(\theta_{1})\cosh{\theta_\ell},\label{app1}
\eeqa
where $\varphi(\theta_{ij})=\varphi(\theta_i-\theta_j)$. Observe that
 this is indeed in agreement with the expression in \eqref{qj}. The expression \eqref{app1} can be derived using the DMEs of $q_i(x,t)$ given by
\beqa
\langle \ltheta|q_i|\rtheta \rangle_{\mathrm{c}}
  &=&
  m\varphi(\theta_{12})\varphi(\theta_{23})\cdots\varphi(\theta_{\ell-1,\ell})
  \nonumber \\
&\times&h_i(\theta_1)\cosh{\theta_\ell}
  + \mathrm{permutations}. \label{ffq}
\eeqa
Similarly, once we evaluate DMEs for the current $j_i(x,t)$, we can construct
 its average ${\tt j}_i$. The expression in \eqref{qj}, that we want to show, will then follow if the
 DMEs of the currents are obtained from those of the densities by the replacement of
 $\cosh{\theta_n}$ with $\sinh{\theta_n}$
\beqa
\langle \ltheta|j_i|\rtheta\rangle_{\mathrm{c}}
  &=&
  m\varphi(\theta_{12})\varphi(\theta_{23})\cdots\varphi(\theta_{\ell-1,\ell})
  \nonumber \\
&\times&h_i(\theta_1)\sinh{\theta_\ell}
  + \mathrm{permutations}. \label{ffj}
\eeqa

Before embarking upon showing it, we elaborate on the definitions of connected and symmetric DMEs. Formally they are given by, respectively, \cite{PT}
\beqa 
\langle \ltheta|\mathcal{O}|\rtheta\rangle_{\mathrm{c}}
&:=&
 F^{\mathrm{c}}_{2\ell}(\mathcal{O},\rtheta) \nonumber \\
 &:=& \mathrm{FP} \lim_{\delta_k \to 0}
 F_{2\ell}(\mathcal{O};\rtheta+\ri\pi+\rdelta,\ltheta),
 \label{ff1} \\
\langle \ltheta|\mathcal{O}|\rtheta\rangle_{\mathrm{s}}
&:=&
 F^{\mathrm{s}}_{2\ell}(\mathcal{O},\rtheta) \nonumber \\
&:=& \lim_{\delta \to 0}
 F_{2\ell}(\mathcal{O};\rtheta+\ri\pi+\delta,\ltheta)
\eeqa
 where FP means ``finite part'' \cite{Mussardo}, $\rdelta=(\delta_1,\cdots,\delta_\ell)$, and the FF
 $F_\ell(\mathcal{O};\rtheta)$ is defined by 
\begin{equation}
 F_\ell(\mathcal{O};\rtheta)=\langle
  \mathrm{vac}|\mathcal{O}(0)|\rtheta\rangle.
\end{equation}
Notice that with a limit such as in \eqref{ff1}, where the parameters $\delta_k$ differ in each
 component, different orders of limits lead to different
 results which may be singular; this is because when $\delta_k \to 0$, the FF (\ref{ff1})
 becomes singular due to kinematic poles. It is in order to circumvent this ambiguity that one defines connected and symmetric FF's. The connected FF is a finite part, which simply prescribes to set to zero terms with singularities in $\delta_k$ \cite{Mussardo}, whereas the symmetric FF is defined
by sending all parameters to zero simultaneously.\par
 It was pointed out in \cite{PT} that any multi-particle symmetric FF can
 be written solely in terms of the connected FFs. For instance, for a
 two-particle state, the connected FF $F_4^{\mathrm{c}}(\mathcal{O};\theta_1,\theta_2)$ and the
 symmetric FF $F_4^{\mathrm{s}}(\mathcal{O};\theta_1,\theta_2)$ satisfy
\beqa
F_4^{\mathrm{c}}(\mathcal{O};\theta_1,\theta_2)
  &=&F_4^{\mathrm{s}}(\mathcal{O};\theta_1,\theta_2)-\varphi(\theta_{12})F_2(\mathcal{O};\theta_1) \nonumber \\
  &-&\varphi(\theta_{21})F_2(\mathcal{O};\theta_2),
\eeqa
where
 $F_2(\mathcal{O};\theta)=F^{\mathrm{c}}_2(\mathcal{O};\theta)=F^{\mathrm{s}}_2(\mathcal{O};\theta)$ (in the case of a single parameter $\delta_1$, there is no singularity, hence no ambiguity). Applying
 this relation to $j_i$, we have 
\beqa
F_4^{\mathrm{c}}(j_i;\theta_1,\theta_2)
  &=&F_4^{\mathrm{s}}(j_i;\theta_1,\theta_2)-\varphi(\theta_{12})F_2(j_i;\theta_1) \nonumber \\
  &-&\varphi(\theta_{21})F_2(j_i;\theta_2).\label{ff2j}
\eeqa
This can be expressed in terms of FFs of the density $q_i$ thanks to the
 conservation law $\p_tq_i+\p_xj_i=0$, which entails 
\begin{equation}
 F^{\mathrm{s}}_{2\ell}(j_i;\rtheta)=
  \frac{\sum_{k=1}^\ell\sinh{\theta_k}}{\sum_{k=1}^\ell\cosh{\theta_k}}F^{\mathrm{s}}_{2\ell}(q_i;\rtheta). \label{ffqj}
\end{equation}
Hence putting (\ref{ffqj}) into (\ref{ff2j}) yields

\begin{equation}
 F_4^{\mathrm{c}}(j_i;\theta_1,\theta_2)=m\varphi(\theta_{12})h_i(\theta_1)\sinh{\theta_2}+
  \{\theta_1 \leftrightarrow \theta_2\},
\end{equation}
which is consistent with (\ref{ffj}). It is readily seen that for
 multi-particle states similar arguments hold, and thus we obtain
 (\ref{ffj}). Finally the generalized LM formula for the current gives
\beqa
{\tt j}_i&=&m\sum_{\ell=0}^{\infty}\biggl(\prod_{k=1}^{\ell}\frac{\dd
  \theta_k}{2\pi}n(\theta_k)\biggr) \nonumber \\
         &\times& \varphi(\theta_{12})\cdots
  \varphi(\theta_{\ell-1,\ell})h_i(\theta_{1})\sinh{\theta_\ell}.
\eeqa
This exactly coincides with \eqref{qj}. Similar arguments give rise to current averages associated to flows $\ri [Q_k,q_i] + \p_x j_i^{(k)}=0$ with respect to any local conserved quantity $Q_k$ (with odd spin):
\beq
	{\tt j}_i^{(k)} = \int \frc{\dd h_k(\theta)}{2\pi} n(\theta)
	h_i^{\rm dr}(\theta).
\eeq
A full derivation will be given in \cite{CDYfuture}.

\end{document}